# A von Neumann Entropy Measure of Entanglement Transfer in a Double Jaynes-Cummings Model

Samina S. Masood*, Allen Miller†

masood@uhcl.edu, allenmil@syr.edu

*Department of Physics, University of Houston Clear Lake, Houston TX 77058,

†Department of Physics, Syracuse University, Syracuse, NY 13244-1130

*Abstract:* We study the entanglement in a system consisting of two non-interacting atoms located in separate cavities, both in their ground states. A single incoming photon has a non-zero probability of entering either of the two cavities. The Jaynes-Cummings interaction in the rotating wave approximation describes the coupling of each atom with the radiation field. We compute and analyze the atom-atom entanglement, the entanglement between the two photon modes, and also the entanglement between each atom and each photon mode. The measure of entanglement is the von Neumann entropy. For the case in which the two atom-photon systems have identical properties, but allowing for non-resonant conditions, the sum of the atom-atom and photon-modes-entanglement is time independent. The effect of detuning is to decrease the strength of the largest entanglement achieved and to shorten the time for it to occur. The results support the fact that the state of the photons after emergence from cavities is entangled, notwithstanding its single-particle nature. In addition, for the case of resonance and identical cavity parameters, we demonstrate that von Neumann entropy is always greater than or equal to the measure of entanglement known as negativity.

## I. INTRODUCTION

A model system that has recently sparked several studies of atom-atom and photon modes entanglement is that in which two atoms are located in separate cavities. A single photon is initially in a prepared state devised so that it may reach either cavity (Fig. 1). The initial photon has probability $\sin^2\alpha$ for entering cavity A and $\cos^2\alpha$ for entering cavity B. The atoms of both cavities are initially in their ground states $|\downarrow\rangle_j$ with j = A or B. The degree of entanglement of the photon state has bearing on proposals to use this photon state or a similar state for teleportation [1], or for quantum cryptography [2]. There are several papers [3] in the literature that use von Neumann entropy as a measure to study the entanglement of squeezed states in separate cavities to apply it to information systems. In the present study, we use von Neumann entropy to examine entanglement using Jaynes-Cummings (JC) model [4] for two-atom and two photon system in the rotating wave approximation. The atom-photon coupling is described by the Jaynes-Cummings (JC) Hamiltonian.

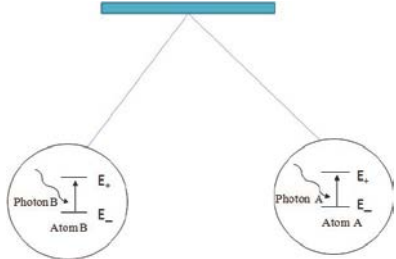

FIG. 1: *Illustration of the double Jaynes-Cummings model. An incident photon (not shown in this Figure) interacts with the atom in cavity A with probability sin2α or cavity B with probability cos2α. The ground state and the excited state energies are designated as E+ and E−, respectively. The figure applies to the case in which the two atoms have identical energy levels.*

In this model, the photon can enable transitions between two states of the atom; these states can be chosen to be the ground state and the first excited state. Pawlowski and Czachor [5] have used a model with two identical systems and photon energy equal to the energy difference between the ground state and the excited state (i.e., resonance conditions assumed). The authors derive the time dependence of the system state and conclude that the existence of entanglement depends on the choice of representation for the quantization of the photon field. A different viewpoint has been presented by van Enck [6]. Arguments are given that the photon state of reference [5] must be entangled. This reasoning is based on consideration of the time development of the system and the resulting entanglement of the pair of atoms. An earlier investigation [7] of this model derives the time development of the state vector for the case of identical parameters for the two cavities, but dealing with both the resonant and non-resonant cases. The initial condition is that both of the two atoms are in their ground-state, while each of the two photon modes have a non-zero probability of containing one photon, with the other mode unoccupied.

Yonac and Eberly [8] have published a study of the double Jaynes-Cummings model, choosing the case in which the properties of the two systems are the same. They consider the initial condition in which both of the two photon modes are unoccupied, with one of the two atoms having a non-zero probability of residing in an excited state, while the other atom is in its ground state. The time dependence of the system state is found, as well as the entanglement, using Wooter's concurrence [9] as the entanglement measure. In references [7] and [8], the atom-atom entanglement is found to have a periodic dependence on time. In addition, reference [8] considers a second initial state which is an admixture of a state with both atoms excited with a state for which both atoms are in their ground state. The initial photon state is the vacuum. For this case, they find intervals of time in which the entanglement is zero (referred to as ``sudden death"). For both initial conditions, a constant-of-the-motion is found that is a linear combination of the six possible concurrences obtained by differing partitions of the variables of the system. Sainz and Bjork [10] have generalized the work of [7] and [8] to the case in which the two systems may have different values for their numerical parameters. They find a new entanglement invariant that is a linear combination of the various wedge products obtained from partitions of the system.

A study of the double Jaynes-Cummings model for the case of resonance is given by Cavalcanti, et al [11]. Different coupling constants for the photon-atom interactions of the two cavities are allowed and negativity is the entanglement measure. The first of the two initial conditions used in reference [8] is considered. The authors draw a phase diagram of atom-atom entanglement versus atom-atom energy. The allowed points in the phase diagram are found to be restricted to a confined area; this area collapses to a curve when the two coupling constants are identical. A partial motivation for the present paper is a resolution of the differing viewpoints expressed in references [5] and [6]. For this reason, we employ an initial state described in the first paragraph of this Introduction. A second motivation is our desire to obtain explicit analytic results for atom-atom and photon-mode entanglement using von Neumann entropy as the measure. Earlier work on the model has used other entanglement measures. In the next Sub-section, we describe our model system in detail. General results for the time evolution of the system state are found in Section II. A calculation of the time dependence of the strength of the entanglement of various sub-systems of the model is given in Section III. A Comparison of double Jaynes-Cummings model with earlier work is in Section IV. The results of the present work are summarized and discussed in Section V.

### A-*Description of the Model System*

The atoms in each of the two cavities are labeled A and B; the same labeling is used for the photon mode that enters cavity A or B, respectively. The Hamiltonian of the Jaynes-Cummings model in the rotating wave approximation [4] can then be used to describe atom j (as j = A, B) and its interaction with photon mode j:

$$H_j = \hbar\omega_j(a_j^\dagger a_j + \frac{1}{2}) + \frac{1}{2}E_j \sigma_{zj} + \hbar g_j(a_j^+ \sigma_{-j} + a_j \sigma_{+j}). \qquad (I-1)$$

The first term is the unperturbed energy of the photon mode labeled j. The operators' $a_j$ and $a^+_j$ are, respectively, the destruction and creation operators for photon mode j. The frequency of photon with mode j is $\omega_j$. The second term gives the unperturbed atomic energies. The magnitude of the atomic energy difference is $E_j$. The x, y, and z components of the matrix $\sigma_j$ are the three Pauli spin matrices; they act on an atomic state exactly as prescribed on a spin one-half particle. The lower row of the spinor corresponds to the ground state, while the upper row corresponds to the excited state. The zero energy of the atomic states is selected halfway between the ground state energy and that of the excited state. The third term is the atom-photon interaction. We adopt the usual definitions of the raising and lowering operators: $\sigma_{+j}= \sigma_{xj} + i\sigma_{yj}$ and $\sigma_{-j} = \sigma_{xj} - i\sigma_{yj}$. This term allows transitions between the two states of each atom, with a corresponding emission or absorption of a photon. The constants 6 $hg_j$ couple atom j to radiation mode j. By an appropriate choice of the phases of the atomic wave functions [4], the constants $g_j$ can be chosen to be real and positive. The complete Hamiltonian is

$$H = H_A + H_B. \qquad (I-2)$$

For the stationary states of $H_j$, Eq.(I-1), we employ the notation and solutions of Hussin and Nieto [12].

## II. TIME EVOLUTION

Our interest will focus on the initial state in which the photon has been prepared in either mode A or mode B and has entered one of the two cavities. Both atoms are in their ground states. The initial state is

$$|\Psi(t = 0) >= \cos\alpha| \downarrow\downarrow 01 > +\sin\alpha| \downarrow\downarrow 10 > . \qquad (II-1)$$

The notation $|\sigma_A, \sigma_B, n_A, n_B >$ refers to the state in which atom j is either in its ground state ($|\sigma_j, \downarrow>$) or is in its upstate ($|\sigma_j, \uparrow>$), for j = A or B. In addition, it prescribes that photon mode j has $n_j$ number of photons. The probabilities that the photon is of mode A or B are given by sin $\alpha$ and cos $\alpha$, respectively (see Fig.1).

### A. General Results for the Time Development

The time development of the initial system state, Eq.(II-1), can now be found:

$$|\Psi(t)) >= exp(-iHt/\ \hbar)|\Psi(t=0) >= \sum_{m=1}^{4} exp(-iE_m t/\ \hbar) < \Psi_m|\Psi(t=0) > |\Psi_m > . \qquad (II-2)$$

Evaluating $< \Psi_m|\Psi(t=0) >$ from Eq.(II-1) and then expanding $|\Psi_m >$ in the standard basis yields the result

$$|\Psi(t) >= r_1(t)| \uparrow\downarrow 00 > +r_2(t)| \downarrow\uparrow 00 > +r_3(t)| \downarrow\downarrow 10 > +r_4(t)| \downarrow\downarrow 01 > . \qquad (II-3)$$

The four time-dependent coefficients $r_s$ (s = 1 to 4) are given

$$r_1(t) = (1/2)(\sin\alpha)\sin2\theta_A[\exp(-iE_4 t/\ 6\ h) - \exp(-iE_2 t/\ 6\ h)],$$
$$r_2(t) = (1/2)(\cos\alpha)\sin2\theta_B[\exp(-iE_3 t/\ 6\ h) - \exp(-iE_1 t/\ 6\ h)],$$
$$r_3(t) = \sin\alpha[\exp(-iE_2 t/\ 6\ h)\sin^2\theta_A + \exp(-iE_4 t/\ 6h)\cos2\theta_A],$$
$$r_4(t) = \cos\alpha[\exp(-iE_1 t/\ 6\ h)\sin^2\theta_B + \exp(-iE_3 t/\ 6\ h)\cos2\theta_B]. \qquad (II-4)$$

In Eq. (11-4), we have defined the angles $\theta_j$ (using j = A, B) by the relations

$$\cos\theta_j = [(\frac{1}{2}) + (\frac{\epsilon_i}{4q_i})]^{1/2} \qquad \sin\theta_j = [(\frac{1}{2}) - (\frac{\epsilon_j}{4q_j})]^{1/2} \qquad (II-5)$$

In Eq.(II-5), we have defined $q_j$ by

$$q_j = [\lambda_i + \frac{\epsilon_j}{4}] \qquad (II-6)$$

where $\lambda_j$ is defined by

$$\lambda_j = \frac{g_j}{E_j} \qquad (II-7)$$

Finally, in Eq. (II-2), we have defined the four stationarystate energies $E_m$ of the states $|\Psi_m >$, (m = 1, 2, 3, 4), are given by

$$E_1 = E_-(A,B), \qquad E_2 = E_-(B,A), \qquad E_3 = E_+(A,B), \qquad E_4 = E_+(B,A). \qquad (II-8)$$

After a straightforward calculation, we obtain the result

$$E_\pm(A, B) = \frac{\varepsilon_A E_A}{2} + (1 + \varepsilon_B)E_B \pm q_{1B} E_B \qquad (II-9)$$

The energies $E_m$ in Eq.(II-8) are depicted in Figure 2, assuming identical parameters for the two cavities and letting $E_{atom} = E_A = E_B$ be the difference in atomic energy of the two states.

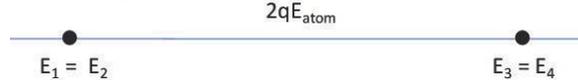

FIG. 2: *Energy values $E_m$ for the stationary states for the case in which the parameters are identical for the two cavities. For this case, $E_1 = E_2$ and $E_3 = E_4$, with $E_3 - E_1 = 2qE_{atom}$ and $q = [(\varepsilon^2/4 + \lambda^2)]^{1/2}$. Thus, the off-resonance parameter $\varepsilon$, and the atom-photon coupling parameter $\lambda$ both contribute to removing the degeneracy between the energy values.*

The probability that atom A (or atom B) is excited is given by $P_A = |r_1|^2$ (or $P_B = |r_2|^2$). Letting $\Lambda_j = q_j E_j/\ 6\ h$, we obtain the result

$$P_A = \frac{\lambda_A^2}{q_A^2}(\sin^2 \alpha) \sin^2 \Lambda_A t, \qquad P_B = \frac{\lambda_B^2}{q_B^2}(\cos^2 \alpha) \sin^2 \Lambda_B t. \qquad (II-10)$$

Henceforth, it is convenient to distinguish atom states from the photon modes. We now let the symbols a and b denote the photon modes that enter cavities A and B, respectively. The designation of the two atoms in cavities A and B remains as A, and B, respectively. Then, the probability $P_a$ (or $P_b$) that photon mode a (or b) is occupied by one photon is $P_a = |r_3|^2$ (or $P_b = |r_4|^2$), with

$$P_a = \sin^2 \alpha [1 - \frac{\lambda_A^2}{q_A^2} \sin^2 \Lambda_A t], \qquad P_b = \cos^2 \alpha [1 - \frac{\lambda_B^2}{q_B^2} \sin^2 \Lambda_B t]. \qquad (II-11)$$

### B. Identical Cavities

For the special case of *identical* atom-photon systems A and B, we have the relations $E_A = E_B = E_{atom}$, $g_A = g_B = g$ and $\omega_A = \omega_B = \omega$. Also, $\varepsilon_A = \varepsilon_B = \varepsilon$, $\lambda_A = \lambda_B = \lambda$, $q_A = q_B = q$, and $\theta_A = \theta_B = \theta$. Further, $E_1 = E_2$ and $E_3 = E_4$. Then, defining $\Lambda = qE_{atom}/6h$, the four probabilities given in Eq.(II-10) and Eq.(II-11) are simplified as:

$$P_A = (\lambda/q)^2 (sin^2\alpha) \sin^2 \Lambda t, \qquad P_B = (\lambda/q)^2 (\cos^2 \alpha) \sin^2 \Lambda t, \qquad (II-12)$$

$$P_a = \sin^2 \alpha [1 - \left(\frac{\lambda}{q}\right)^2 \sin^2 \Lambda t], \qquad P_b = \cos^2 \alpha [1 - \left(\frac{\lambda}{q}\right)^2 \sin^2 \Lambda t] \qquad (II-13)$$

Hence, the probability that atom A [or atom B] is excited rises from zero at time t = 0 to a maximum value of $\left(\frac{\lambda}{q}\right)^2 (\sin^2 \alpha$ , [or $\left(\frac{\lambda}{q}\right)^2 \cos^2\alpha$] at time $t^{\max} = \frac{\pi}{2\Lambda}$. At $t_{max}$, the atoms have their largest value for entanglement, since excitation of A requires that B be in its ground state and vice versa.

The probability that neither atom is excited is $|r_3|^2 + |r_4|^2 = \cos^2(\Lambda t) + (\frac{\varepsilon^2}{4q^2}) \sin^2 \Lambda t$ and has its lowest value of $\frac{\varepsilon^2}{4q^2}$ when t = $t_{max}$.

### C. Identical Cavities and Resonance

If, also, resonance occurs, (6 $h\omega$ = $E_{atom}$, $\varepsilon$ =0, $\theta$= $\pi$/4, q =$\lambda$ =g /$E_{atom}$), Eq.(II-4) simplifies to Eq. (*II-14*). Note that for resonance, the interaction strength g replaces $\Lambda$, since $\Lambda$ = $qE_{atom}$/ 6 h at resonance simplifies to $\Lambda$ = *g*.

$$r_1(t) = -i(\sin\alpha)\exp(-i\omega t)\sin gt,$$
$$r_2(t) = -i(\cos\alpha)\exp(-i\omega t)\sin gt,$$
$$r_3(t) = (\sin\alpha)\exp(-i\omega t)\cos gt,$$
$$r_4(t) = (\cos\alpha)\exp(-i\omega t)\cos gt. \qquad (II-14)$$

At time $t_{max,res}$ = $\pi$/(2*g*), there is maximum probability for atom excitation. It can be noted that the effect of detuning is to shorten the period of oscillation, since $t_{max} \leq t_{max,res}$.

## III. ENTANGLEMENT

To evaluate the atom-atom entanglement using von Neumann entropy as the measure, Eq.(II-3) indicates that it is necessary to consider three atomic states separately, as the photon states lie in a three-dimensional vector space generated by photons..

### A. General Results

The simplest basis of the three-dimensional vector space of the photon states is

$$|\Omega_0> = |0>_a \otimes |0>_b, \qquad |\Omega_1> = |0>_a \otimes |1>_b, \qquad |\Omega_2> = |1>_a \otimes |0>_b. \qquad (III-1)$$

The state $|\Psi(t)>$ can be written as a linear combination of these three photon states:

$$|\Psi(t)> = \sum_{k=0}^{2} \beta_k(t)|h_k(t)> \otimes |\Omega_k>. \qquad (III-2)$$

The three normalized functions $|h_k(t)>$ of the variables of atoms A and B are

$$|h_0(t)> = N_{AB}^{1/2}[r_1(t)|\uparrow\downarrow> + r_2(t)|\downarrow\uparrow>], \qquad |h_1> = |h_2> = |\downarrow\downarrow>, \qquad (III-3)$$

with the notation $|\sigma_A\sigma_B> = |\sigma_A> \otimes |\sigma_B>$. The normalization constant $N_{AB}$ is

$$N_{AB} = [|r_1|^2 + |r_2|^2]^{-1}. \qquad (III-4)$$

The constants $\beta_k(t)$ give the probability $|\beta_k(t)|^2$ that a measurement of the system state at time t will yield the result $|h_k> \otimes |\Omega_k>$. They are given by

$$\beta_0(t) = N_{AB}^{-1/2}, \qquad \beta_1(t) = r_1(t), \qquad \beta_2(t) = r_2(t). \qquad (III-5)$$

We denote the von Neumann entropy of the pair of atoms [13] by $S_{A/B}$. Since there are three atomic functions $|h_k>$, the atom-atom entropy $S_{A/B}(k,t)$ of each is computed separately, and then weighted according to the probabilities $|\beta_k(t)|^2$. This yields the net atom-atom entropy $S_{A/B}$. The Schmidt Decomposition [14] of the three atomic wave functions $|h_k>$ is

$$|h_k> = \sum_{s=1}^{2} \sqrt{p_s(k)} |h_{k,s}>_A \otimes |h_{k,s}>_B. \qquad (III-6)$$

In Eq.(III-6), the pair of states $|h_{k,s}>_j$ denotes (as s = 1, 2) an orthonormal basis for the states of atom j. The positive constants $pp_s(k)$ weight the contributions of each member of the basis. For k = 0, a choice for the basis for the states of atom j, and for the coefficients $p_s(k)$ is

$$|h_{0,1}>_A = \frac{r_1}{|r_1|}|\uparrow>_A, \qquad |h_{0,2}>_A = \frac{r_2}{|r_2|}|\downarrow>_A,$$

$$|h_{0,1}>_B = \frac{r_1}{|r_1|}|\downarrow>_B, \qquad |h_{0,2}>_B = \frac{r_2}{|r_2|}|\uparrow>_B$$

$$p_1(0) = N_{AB}|r_1|^2, \qquad p_2(0) = N_{AB}|r_2|^2. \qquad (III-7)$$

or k = 1 and 2, we can choose

$$|h_{k,1}>_j = |\downarrow>_j, \qquad |h_{k,2}>_j = |\uparrow>_j; \qquad j = A, B,$$

$$p_1(k) = 1, \qquad p_2(k) = 0. \qquad (III-8)$$

The von Neumann entropy [13] of atomic state $|h_k>$ is (logarithms with base 2)

$$S_{A/B}(k) = -\sum_{s=1}^{2} p_s(k) \log_2 p_s(k). \qquad (III-9)$$

The desired atom-atom entropy is the weighted average

$$S_{A/B}(t) = \sum_{s=1}^{2} |\beta_k(t)|^2 S_{A/B}(k,t), \qquad (III-10)$$

and can be evaluated from Eqs.(III-5), (III-7), (III-8) and (III-9), yielding our result for the atom-atom entropy:

$$S_{A/B}(t) = -P_A \log_2\left[\frac{P_A}{(P_A + P_B)}\right] - P_B \log_2\left[\frac{P_B}{(P_A + P_B)}\right]. \qquad (III-11)$$

Now consider the entanglement $S_{a/b}$ between the two *photon modes* a and b, corresponding to cavities A and B, respectively. To compute $S_{a/b}$ between the two modes, it is only necessary to repeat the calculations of Eqs.(III-1) to (III-11), reversing the roles of the photon states and the atomic states. The results for $S_{a/b}$ and also for the atom-photon entropies $S_{A/a}$, $S_{A/b}$, $S_{B/a}$ and $S_{B/b}$ are given by a generalization of Eq.(III-11):

$$S_{\mu/\xi} = -P_\mu \log_2[P_\mu/(P_\mu + P_\xi)] - P_\xi \log_2[P_\xi/(P_\mu + P_\xi)]. \qquad (III-12)$$

Here, the pairs $(\mu,\xi)$ take on the six choices (A, B), (a,b), (A,a), (A,b), (B,b) and (B,a) and thereby generate the six von Neumann entropies desired. Note that Eqs.(II-10) and (II-11) give all probabilities on the right side of Eqs.(III-11) and (III-12).

## B. *Identical Cavities*

If we again specialize to the case when Eqs.(II-10) to (II-13) are valid, Eq.(III-12) yields the following results for $S_{A/B}(t)$ and $S_{a/b}(t)$:

$$S_{A/B}(t) = G(\alpha)(\lambda/q)^2 \sin^2 \Lambda t, \qquad S_{a/b}(t) = G(\alpha)[1 - (\lambda/q)^2 \sin^2 \Lambda t], \qquad (III-13)$$

$$G(\alpha) = -(\sin^2 \alpha)\log_2(\sin^2 \alpha) - (\cos^2 \alpha)\log_2(\cos^2 \alpha). \qquad (III-14)$$

The function $G(\alpha)$ is plotted in Fig. 3. It can be regarded as the photon-entry function as it is the pre-factor for the entropies in Eq.(III-13) and it represents the only dependence on $\alpha$ of the entropies. The sum of the atom-atom and photon mode entanglements is time-independent, for the case where Eqs.(II-13)) and (III-13) are valid (i.e., identical parameters for cavities A and B):

$$S_{A/B}(t) + S_{a/b}(t) = G(\alpha). \qquad (III-15)$$

It can be remarked in Fig. 3 that when the probabilities are equal for the photon entering cavities a or b, then $\alpha = \pi/4$ and the sum in Eq.(III-14) is at its maximum value of $G(\pi/4)= 1$. We use normalized function $G(\alpha)$ in the remaining text later to extract the dependence of entropy on these parameters G and $\alpha$ more clearly. For identical cavity parameters, the like atom-photon mode entropies simplify to the results

$$S_{A/a}(t) = -(\sin^2 \alpha) \sum_{i=1}^{2} f_i \log_2 f_i, \qquad (III-16a)$$

with

$$f_1 = f_1(t) = (\lambda/q)^2 \sin^2 \Lambda t, \; f_2 = f_2(t) = 1 - f_1(t),$$

and

$$S_{B/b}(t) = -\cos^2 \alpha \sum_{i=1}^{2} f_i \log_2 f_i. \qquad (III-16b)$$

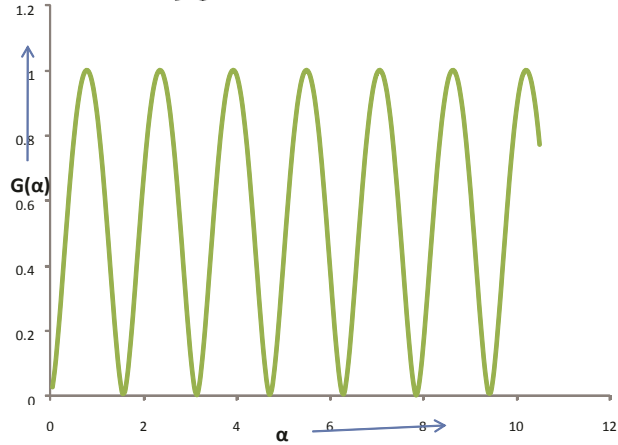

FIG. 3: *Plot of $G(\alpha)$ as a normalized function of $\alpha$ given in Eq.(III-14). $G(\alpha)$ serves as a pre-factor for the atom-atom entropy $S_{A/B}$ and for the photon modes entropy $S_{a/b}$ as can be seen from Eq.(III-13).*

The opposite atom-photon mode entropies are given for this case by

$$S_{A/b}(t) = -(\sin^2 \alpha)f_1 \log_2(b_1) - (\cos^2 \alpha)f_2 \log_2(b_2), \qquad (III-16c)$$

where

$$b_1 = K/(K+1), \qquad b_2 = 1/(1+K), \qquad K = K(\alpha,t) = (\tan^2 \alpha)(f_1/f_2).$$

and

$$S_{B/a}(t) = -(\cos^2 \alpha) f_1 \log_2(b_3) - (\sin^2 \alpha) f_2 \log_2(b_4), \qquad (III-16d)$$

$$b_3 = L/(L+1), \qquad b_4 = 1/(1+L), \qquad L = L(\alpha,t) = (\cot^2 \alpha)(f_1/f_2).$$

If, in addition, we choose $\alpha = \pi/4$, all the atom-photon entropies are equal:

$$S_{A/a}(t) = S_{B/b}(t) = S_{A/b}(t) = S_{B/a}(t) = -\frac{1}{2}\sum_{i=1}^{2} f_i \log_2 f_i . \qquad (III-17)$$

### C. IDENTICAL CAVITIES AND RESONANCE

Specializing further to the resonant case $\varepsilon = 0$, $\Lambda = G$ Eqs.(III-14), (III-16a) and (III-16b) yield, for the choice $\alpha = \pi/4$,

$$S_{A/B}(t) = \sin^2 gt, \qquad S_{a/b}(t) = \cos^2 gt, \qquad (III-18a)$$

$$S_{A/a}(t) = S_{A/b}(t) = S_{B/b}(t) = S_{B/a}(t) = -(1/2)[c\log_2(c) + d\log_2(d)], \qquad (III-18b)$$
$$c = \cos^2 gt, \qquad d = \sin^2 gt.$$

The atom-photon entropies given by Eqs.(III-16a), (III-16b) and (III-17) are all zero at t = 0. For the resonant sub-case, Eqs.(III-18), these entropies are also zero at t = $\pi/(2g)$. In contrast, the atom-atom entropy is zero at t=0, but reaches maximum size at t = $\pi/(2g)$, as can be seen, for example, from Eq.(III-18) for the resonant case. Note that the photon-photon entropy has the opposite behavior, as it must, since the sum on the left side of Eq.(III-15) is constant. It may also be remarked that the maximum value attainable for the four atom-photon entropies given by Eq.(III-17) is 1/2, and occurs at resonance, and when $\alpha = \pi/4$. This value is just one-half of the maximum entropy possible of 1. It occurs at time t = $\pi/(4g)$.

### D. *Graphs of Entropies in Entanglement*

In this section we display some of the results for entanglement derived in Eq.(III-13). All of the plots correspond to the case in which the parameters for the two cavities are identical. It is also worth-mentioning that according to Eq.(III-14), the maximum value of $G(\pi/4) = 1$.

In Figure (4a) we display the graphs for the time evolution of the atom-atom entropy $S_{A/B}(t)$ and the photon modes entropy $S_{a/b}(t)$ for the resonance case ($\varepsilon^2 = 0$) and with the choice $\alpha = \pi/4$ (equal cavity-entry probabilities), employing Eqs.(III-18). The sum $S_{A/B}$ and $S_{a/b}$ maintains at all times the maximum entropy value of 1, as required by Eq.(III-15). The graph shows the oscillatory transfer of entropy back and forth between $S_{A/B}$ and $S_{a/b}$, starting with maximum $S_{a/b}$ at t=0 and reaching maximum $S_{A/B}$ at t=$\pi/(2g)$.

Figure (4b), however, indicates the non-resonant (non-zero $\varepsilon$) behavior of entropies for a special case of $(\lambda/q)^2$=0.5 and G($\alpha$)=0.75.

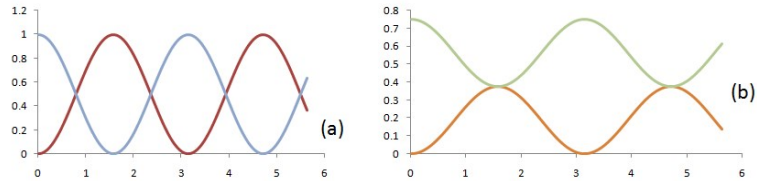

FIG. 4: *Plot of the photon-mode entropy $S_{a/b}$ (blue/green, on-line) and the atom-atom entropy $S_{A/B}$ (red/orange, on-line) as a function of time for a resonant (a) and a non-resonant (b) case. In (a), the horizontal coordinate is gt ($\Lambda = g$ for resonance), and with $G(\alpha)=1$. In (b), the horizontal coordinate is $\Lambda t$ with $\Lambda = qE_{atom}/6 \hbar$. Also $(\lambda/q)^2=0.5$ and G=0.75.*

We show in Figure 5, the dependence of $S_{a/b}(t)$ (part a) and $S_{A/B}$ (part b) on the cavity-entry function $G(\alpha)$, for three different values of $G(\alpha)$. Resonance is assumed. As can be seen from Eq.(III-13), both entropies are proportional to G($\alpha$), and this is their only dependence on the entry parameter $\alpha$. Hence the maxima and minima always occur at the same value of t, regardless of the value of $\alpha$.

Finally we turn to the dependence of the entanglement on the amount of detuning, that is, the case when $\varepsilon$ is non-zero. For this purpose, we shall use the deviation of $(\lambda/q)^2$ from 1 as the measure of detuning, noting that the relation

$$(\lambda/q)^2 = [1 + (\frac{\varepsilon E_{atom}}{2 \hbar g})^2]^{-1}$$

can easily be obtained from a straight forward calculations. The time development of $S_{A/B}$ and $S_{a/b}$ for a few particular values of $(\lambda/q)^2$ is explicitly shown in Figure 6.

Summarizing the features of Figure 6, it can be noted that the sum of the atom-atom entropy $S_{A/B}$ and the photon modes entropy $S_{a/b}$ remains constant, even for the case of detuning, as is required by Eq.(III-15). Second, as $(\lambda/q)^2$ drops from 1, the maximum value of $S_{A/B}$ and the minimum value of $S_{a/b}$ are both attained at odd integral multiples of $\pi\Lambda t/2$ for all values of the detuning parameter.

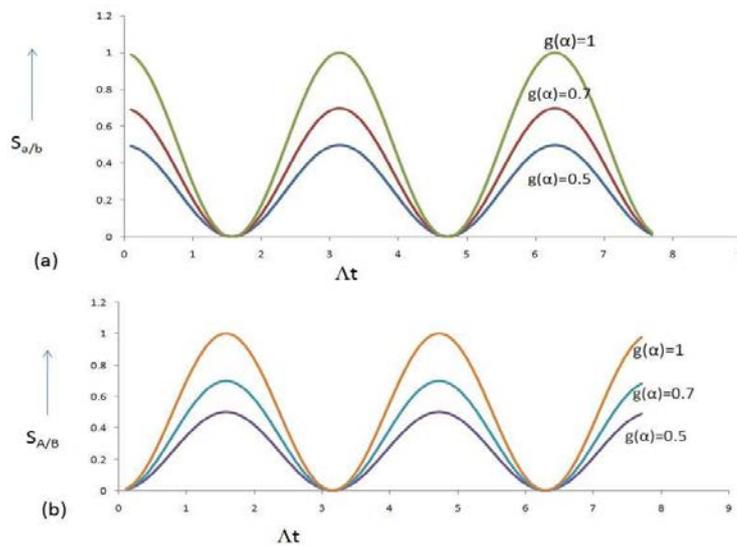

FIG. 5: *Plot of photon modes entanglement $S_{a/b}$ and atom-atom entanglement $S_{A/B}$ versus $\Lambda t$ (where $\Lambda = qE_{atom}/ 6 \hbar$) for three different values of the photon entry function $G(\alpha)$, defined by Eq.(III-13) and indicated as g($\alpha$) in this figure. The dependence of the entanglement on photon probabilities is simply expressed by the multiplicative factor $G(\alpha)$. Resonance is assumed here, so that $\lambda/q=1$ and $\Lambda = g$ in Eq.(III-13).*

## IV.  COMPARISON WITH PREVIOUS WORKS

### A.  VAN ENCK

Van Enck [6] has examined the issue as to whether the two-mode one-photon state is entangled, citing conflicting statements in the literature

$$|\Omega> = \frac{1}{\sqrt{2}}(|01> + |10>) \quad . \quad (IV-1)$$

He gives an argument that this state is, in fact, entangled, based on consideration of an interaction of each photon mode with one of a set of two atoms. The work of reference [5] challenges whether one can say definitively whether the one-photon state Eq.(IV-1) is entangled, arguing that it depends on the representation used to express the state. A second argument appearing in the literature that the Eq. (IV-1) is not entangled is that only a single photon is present. However, Van Enck gives a qualitative argument that the state must be entangled. He considers an initial state identical with our state, Eq. (II-1), (if our $\alpha$ is chosen as $\pi/4$). The interaction between atom and photon in his model corresponds to the third term of Eq.(I-1), applied with resonance conditions. Then, at a later time, there is certainty that the photon is absorbed; one of the two atoms is thereby excited, with the other atom remaining in its ground state.

The system state at this later time is

$$|\Psi> = |h> \otimes |00>, \qquad |h> = \sqrt{1/2}[|\uparrow\downarrow> + |\downarrow\uparrow>]. \qquad (IV-2)$$

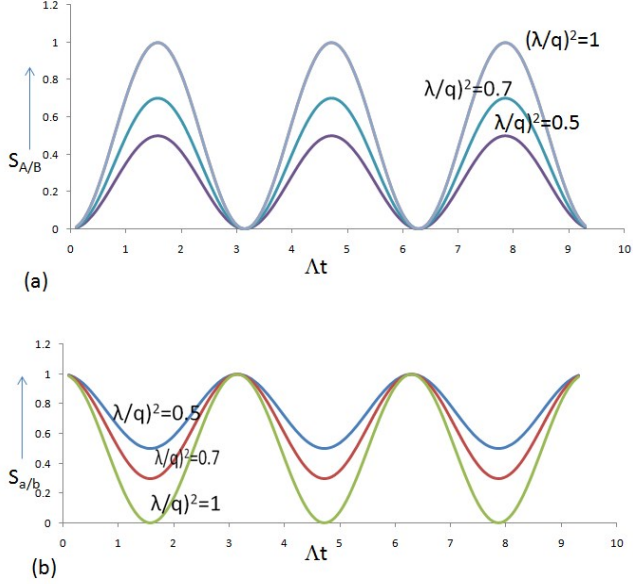

FIG. 6: Plot of the atom-atom entanglement $S_{A/B}$ (part a) and the photon-mode entanglement $S_{a/b}$ (part b) versus $\Lambda t$ for the case of resonance ($\lambda/q = 1$) and for two cases of detuning, i.e., values of 0.5 and 0.7 for $(\lambda/q)^2$. The sum of $S_{A/B}$ and $S_{a/b}$ is time-independent, as displayed in Eq.(III-15). Identical parameters for two cavities are assumed. Also, the graph applies to the case $G(\alpha)=1$.

Now, the atomic state $|h>$ is clearly entangled, as it cannot be written as a product of a function of variables of one atom with that of the other. Since only local operations have generated the entangled state $|h>$ of Eq.(IV-2), it is concluded in reference [6] that the initial state Eq.(IV-1) must also be entangled.

It can be noted that the result of Eq.(II-14) gives quantitative confirmation to this argument. From it, the time for complete transfer of entanglement from photon modes to atoms is $\pi/2g$. (See the discussion below Eq.(II-12).) Further, the result for the entanglement $S_{a/b}(t)$ between photon modes, as given in the first relation of Eqs.(III-18), yields $S_{a/b}(t = 0) = 1$, its maximum value. This confirms quantitatively that the photon state Eq.(IV-1) is maximally entangled.

### B. PAWLOWSKI AND CZACHOR [5]

Here, the double Jaynes-Cummings model is examined for the case of identical atoms, identical coupling constants and resonance. The initial state chosen with $\alpha = \pi/4$.

The authors obtain a result equivalent to Eq.(II-11) for the time-dependence of the probability amplitude that either atom A or atom B is excited. At $t = \pi/2g$, the two atoms are maximally entangled and the system state is given by Eq.(IV-2).

However, the authors then state that it cannot be concluded that the initial photon state, Eq.(IV-1) is entangled. Instead, arguments are given that, for any specified photon state, its degree of entanglement depends on the representation used to describe it. The authors give two examples of representations of the photon state Eq.(IV-1), and claim that any choice of a measure for entanglement will fail to realize a non-zero result.

A resolution of the differing conclusions expressed in [5] and [6] can be found, if it is recognized that the entanglement concept is meaningful only if a clear specification is made of the subsystems experiencing the interconnectedness. Thus, the result for $S_{a/b}$, Eq.(III-18) gives the entanglement between photon modes $a$ and $b$. If the procedures used to obtain $S_{a/b}$ are applied to the representations of our Eq.(II-1) discussed in [5], our results would again be obtained and the photon state would be found to have the maximum entanglement of 1.

The fact that the photon state, Eq.(IV-1) is that of just *one* photon is not relevant. In this state, a measurement revealing that one of the two photon modes is vacant requires that a simultaneous measurement of the other mode yield that it contain one photon. Thus, the two photon modes are entangled.

### C. YONAC, YU AND EBERLY [8]

In this pair of papers, the authors study the double Jaynes-Cummings model for the case of identical atoms, photon frequencies, and coupling constants, but allowing for non-resonant conditions. They examine the evolution of entanglement between the various sub-systems, using concurrence [9] as the measure. The following two initial states are considered:

$$|\Psi_{initial}> = \cos\alpha|\uparrow\downarrow 00> + \sin\alpha|\downarrow\uparrow 00>, \qquad (IV-3a)$$

$$|\Psi_{initial}> = \cos\alpha|\uparrow\uparrow 00> + \sin\alpha|\downarrow\downarrow 00>. \qquad (IV-3b)$$

The Eq.(IV-3a) can be viewed as a transformation of the initial state Eq.(II-1), if the roles of atomic states and photon states are interchanged. The interchange of $|\downarrow>$ with $|0>$ and of $|\uparrow>$ with $|1>$ performs the transformation.

For this initial state, the authors find that the sum $S_{A/B}(t) + S_{a/b}(t)$ is time-independent, parallel to the result Eq.(III-15) obtained in the present study, in which von Neumann entropy is the measure.

For Eq.(IV-3b), the authors find that the entanglement is a non-analytic function of time, experiencing time periods with non-zero duration, in which the entanglement is zero (sudden death). A *time-independent* linear combination of concurrences is found for initial states of the form of the Eq.(IV-3a) but not for those of the second state.

### D. SAINZ AND BJORK [10]

The study of Sainz and Bjork extends the work of [7] and [8] to the case in which the system Aa is not necessarily identical to that of Bb. The time development is found for the two initial states of Eqs.(IV-3). Also, the entanglements corresponding to various partitions of the system are found, using wedge products as the measure. The writers find a linear combination of the wedge products that is time-independent, valid for both choices of the initial state. They note that invariants for a more general class of initial states (e.g., linear combinations of Eqs.(IV-3)) have not been found, but may exist.

In the present study, the sum $S_{A/B}(t) + S_{a/b}(t)$, Eq.(III-15), is invariant only for the case for which the system Aa is identical to that of Bb. It is easy to show that no linear combination of the six entropies of Eq.(III-12) is invariant, if, for example, the atomic energy differences $E_A$ and $E_B$ differ.

### E. CAVALCANTI, ET AL [11]

The authors study the model with the Hamiltonian of Eqs.(I-1) and (I-2), restricting to the resonance situation, but allowing the two coupling constants to differ. *Negativity* is used as the entanglement measure, and the Eq.(IV-3a) serves as the initial state.

It is found that the initial entanglement between the photon modes is fully or partly transferred to the two atoms, with the rate of transfer depending on the ratio of the coupling constants, and on time. A phase diagram is drawn of the entanglement between the atoms versus their energy. As time develops, only a restricted area of this phase space is occupied. For the special case of equal coupling constants, the allowed area in phase space collapses to a curve and the entanglement becomes a single-valued function of the energy.

We compare their results with ours, for the case of identical cavity parameters and resonance. We also assume equal probabilities for photon cavity entry, so that $G(\alpha) = 1$. Let $U_{AB}$ denote the (dimensionless) sum of the energies of atoms A and B, relative to their energy when both atoms are in their ground states. The unit for energy is $E_{atom}$ which is the common difference in energy between the excited and ground states of the atoms. Thus, $U_{AB}$ is defined by $(P_A E_A + P_B E_B)/E_{atom}$.

Then, from Eq.(II-10) to Eq.(II-11), our result is

$$U_{AB} = (P_A E_A + P_B E_B)/E_{atom} = \sin^2 gt. \qquad (IV-4)$$

To obtain Eq.(IV-4), we have assumed that $E_A = E_B = E_{atom}$, as well as $\lambda = q$ (resonance). For this case, it follows that there is a simple equality between $U_{AB}$ and the atom-atom entropy $S_{A/B}$:

$$S_{A/B}(t) = G(\alpha) U_{AB}(t). \qquad (IV-5)$$

However, when systems Aa and Bb differ, Eq.(IV-5) fails. Then, Eq.(III-11) and the equation just before Eq.(II-8) show that the rate of energy gain and entanglement gain are no longer proportional, due to the differing forms for the time dependence of $P_A$ and $P_B$. So, similar to the results of [11], our results for the phase diagram of $S_{A/B}(t)$ versus $U_{AB}(t)$ would show that $S_{A/B}(t)$ is no longer a single valued function of $U_{AB}(t)$, as it is in Eq.(IV-5). The allowed $(S_{A/B}, U_{AB})$ will lie in a bounded area in phase space, as in [11].

Worthy of note is that differing choices for the entanglement measure yield non-equivalent results for its analytic form. As an example, consider the result for the entanglement $N_{A/B}(t)$ (termed the negativity) between the atoms computed in [11] for the case of identical Aa, Bb systems. In our notation, [15], their result is

$$N_{A/B}(t) = [(U_{AB}(t) - 1)^2 + U_{AB}^2(t)]^{1/2} + U_{AB}(t) - 1. \qquad (IV-6)$$

Eq.(IV-6) is applicable for the case of resonance, identical cavity parameters, and maximum initial entanglement of the photon modes ($\alpha = \pi/4$).

The two results Eqs.(IV-5) and (IV-6) express the time-development of the entanglement for the same initial state Eq.(II-1). The former result uses von Neumann entropy and the latter uses negativity. Both entanglements are zero at time t = 0 and have their maximum value of 1 at time t = $t_{max,res} = \pi/(2g)$. They are thereby in qualitative agreement with the meaning of entanglement.

However at all intermediate times the von Neumann entropy exceeds the negativity, as can be shown by a straightforward comparison between the right-hand-sides of Eqs.(IV-5) and (IV-6). That is,

$$S_{A/B} > N_{A/B}. \qquad (IV-7)$$

for $0 < t < \pi/(4g)$.

This comparison of our results with the existing results show that our results are important as we give the timedependence of a pure state of a double Jaynes-Cummings model (using the rotating wave approximation) in Eqs.(II-2) and (II-3) without putting any further conditions on the states themselves. The initial state represents a photon emitting from one of the laser cavities A and B with arbitrary probabilities for entering one of two cavities, each containing one atom, with the two atoms not necessarily identical. The entanglements between any two of the four components of the system are found and are given in Eqs.(II-11),(II-12) and (III-12), with von Neumann entropy as the entanglement measure. A quantitative analysis shows the time dependence of the entanglement, as well as its dependence on the cavity entry function $G(\alpha)$, defined by Eq.(III-13). The behavior of atom-atom entropy and the photon-photon entropy is discussed in detail in Section III. In the next section we summarize and discuss the unique and interesting features of these results.

## V. RESULTS AND DISCUSSION

We explicitly plot the atom-atom $S_{A/B}$ and photon-mode entropy $S_{a/b}$ to show their dependence on the angle $\alpha$ and the time t. For the case in which each of the two atom-photon systems (Aa and Bb) have identical properties, the sum $S_{A/B} + S_{a/b}$ of the atom-atom and photon-mode entanglements is a constant-of-the-motion, as displayed in Eq.(III-15). For this case, the atom entanglement $S_{A/B}$ increases from its initial zero value, reaching its maximum value at time $t_{max} = \pi/(2\Lambda)$, where $\Lambda$ is defined before Eq.(II-10). This increase is at the expense of the entanglement between the two photon modes; the latter entanglement declines and reaches a minimum value at $t_{max}$. For the case of resonance and equal cavity-entry probabilities, the atoms become fully entangled at this time, so that $S_{A/B} = 1$.

If the two systems Aa and Bb do not have identical properties, then the nature of entanglement swapping is not fully understood. It is no longer true that $S_{A/B} + S_{a/b}$ is time-independent. In fact, for von Neumann entropy as the measure, it can be shown that there is no linear combination of the six entropies of Eq.(III-12) that is time-independent.

In Section IV, Parts A and B, we discuss differences in opinion as to whether the photon state Eq.(IV-1), is entangled. Arguments are given for an affirmative answer.

Connections between the results of this study with those of references [8], [10] and [11] are given in Section IV, Parts C, D, and E respectively. These references employ measures for entanglement differing from von Neumann entropy. While all measures for entanglement share certain basic qualitative features, they do not yield results of the same form. For example, compare Eqs.(IV-5) and (IV-6). As noted at the end of Section IV, use of the negativity as the measure yields a smaller value for the entanglement than does use of von Neumann entropy for all times between minimum and maximum values of the entanglement. Work that illuminates the meaning of the differences between the various measures would be of value, as results for entanglement transfer depend on the measure used. An interesting open question is whether there exists a sum of entanglements of sub-systems of the double Jaynes Cummings model which is a constant of the motion for all values of the parameters of the model.

The Jaynes-Cummings model has been recently used to study entanglement in a system of four atoms [16]. This study of a double Tavis-Cummings model (DTCM) is developed to simulate the entanglement dynamics of realistic quantum information processing where two entangled atom-pairs *AB* and *CD* are distributed in such a way that atoms *AC* are embedded in a cavity *a* while *BD* are located in another remote cavity *b*. The methods used in this analysis can be extended to study such systems, also.

In the work of reference [17], the entanglement invariant obtained by Sainz and Bjork [10] is discussed. The authors [17] present arguments that this invariant is related to linear entropy between two independent systems. It has been demonstrated experimentally [18] that the displacement measurement of a light beam below the standard quantum limit can be obtained using continuous wave superposition of spatial modes. In such experiments the multimode squeezed light is obtained by mixing a vacuum squeezed beam and a coherent beam that are spatially orthogonal to each other. Although the resultant beam is not squeezed, it is shown to have strong internal spatial correlations. The work of Ref. [18] shows that the position of such a light beam can be measured using a split detector with an increased precision compared to a classical beam. This method can be used to improve the sensitivity of small displacement measurements. We conclude by citing three references[16-18] that deal with models similar to the model discussed here.